\documentclass[12pt, a4paper]{article}

\usepackage[utf8]{inputenc}
\usepackage{graphicx}
\usepackage{url}
\usepackage{amsmath}
\usepackage{amssymb}
\usepackage{dsfont}

\usepackage{ifpdf}
\ifpdf
\usepackage[%
  pdftitle={Statistical boosting review},%
  pdfauthor={Andreas Mayr, Benjamin Hofner, Elisabeth Waldmann, Tobias Hepp, Olaf Gefeller, Matthias Schmid},%
  pdfstartview=FitH,%
  bookmarks=false,%
  bookmarksopen=false,%
  breaklinks=true,%
  colorlinks=true,%
  linkcolor=black,anchorcolor=black,%
  citecolor=black,filecolor=black,%
  menucolor=black,pagecolor=black,%
  urlcolor=black]{hyperref}
\else
\usepackage[%
  breaklinks=true,%
  colorlinks=true,%
  linkcolor=blue,anchorcolor=blue,%
  citecolor=blue,filecolor=blue,%
  menucolor=blue,pagecolor=blue,%
  urlcolor=blue]{hyperref}
\fi

\newcommand{\titleman}[1]{\begin{center}{\bf \LARGE
#1}\end{center}}

\newcommand{\bm}[1]{\mbox{\boldmath{$#1$}}}

\long\def\symbolfootnote[#1]#2{\begingroup%
    \def\thefootnote{\fnsymbol{footnote}}\footnote[#1]{#2}\endgroup}

\usepackage{float}

\usepackage{xcolor}
\usepackage[normalem]{ulem}

\makeatletter
\newcommand\floatc@simplerule[2]{{\@fs@cfont #1} #2\par}
\newcommand\fs@simplerule{\def\@fs@cfont{\bfseries}\let\@fs@capt\floatc@simplerule
  \def\@fs@pre{\hrule height1.2pt depth0pt \kern4pt}%
  \def\@fs@mid{\vspace*{0.5em} \hrule height.3pt depth0pt \vspace*{0.8em} \kern4pt}%
  \def\@fs@post{\kern4pt \hrule height1.2pt depth0pt \kern4pt \relax}%
  \let\@fs@iftopcapt\iftrue}
\makeatother

\floatstyle{boxed}
\newfloat{algorithm}{thp}{alg}	
\floatname{algorithm}{Box}

\begin{document}

\clearpage

\titleman{An update on statistical boosting in biomedicine }

\begin{center}
Andreas Mayr\symbolfootnote[1]{ \textit{Address for correspondence:}
Andreas Mayr, Institut f\"ur Medizininformatik, Biometrie und
Epidemiologie, Friedrich-Alexander Universit\"at
Erlangen-N\"urnberg, Waldstr. 6, 91054 Erlangen.\\ Email:
\href{mailto:andreas.mayr$at$fau.de}{andreas.mayr@fau.de}}$^{1,3}$,
Benjamin Hofner$^2$, Elisabeth Waldmann$^1$,\\ Tobias Hepp$^1$, Olaf Gefeller$^1$, Matthias Schmid$^{3}$ \vspace{0.5cm}

\begin{footnotesize}
 \footnotemark[1] Institut f\"ur Medizininformatik, Biometrie und Epidemiologie,\\
Friedrich-Alexander-Universit\"at Erlangen-N\"urnberg, Germany\\
\footnotemark[2] Section Biostatistics,
Paul-Ehrlich-Institut, Langen, Germany \\
\footnotemark[3] Institut f\"ur medizinische Biometrie, Informatik und Epidemiologie, \\
Rheinische Friedrich-Wilhelms-Universit\"at Bonn, Germany
\end{footnotesize}

\end{center}
\normalsize \vspace{1cm}

\begin{abstract}

Statistical boosting algorithms have triggered a lot of research during the last decade. They combine a powerful machine-learning approach with classical statistical modelling, offering various practical advantages like automated variable selection and implicit regularization of effect estimates. They are extremely flexible, as the underlying base-learners (regression functions defining the type of effect for the explanatory variables) can be combined with any kind of loss function (target function to be optimized, defining the type of regression setting). In this review article, we highlight the most recent methodological developments on statistical boosting regarding variable selection, functional regression and advanced time-to-event modelling. Additionally, we provide a short overview on relevant applications of statistical boosting in biomedicine.

\end{abstract}

\section{Introduction}

Statistical boosting algorithms are one of the advanced methods in the toolbox of a modern statistician or data scientist \cite{mayr2014evolution}. They offer multiple advantages in the presence of high-dimensional data as they can deal with more potential candidate variables than observations ($p > n$ situations) while still yielding classical statistical models with well-known interpretability \cite{BuhlmannHothorn06, TutzBinder}. Key features in this context are automated variable selection and model choice \cite{Hofner:unbiased:2011,Kneib2009}.

The field of research is methodologically situated between the world of statistics and computer science. They bridge the gap between two rather different point of views on how to gather information from data \cite{breiman2001statistical}: on the one hand, there is the classical statistical modelling community that focuses on models \textit{describing} and \textit{explaining} the outcome in order to find an approximation to the underlying stochastic data generation process. On the other hand, there is the machine learning community that focuses primarily on algorithmic models \textit{predicting} the outcome while treating the nature of the underlying process as unknown. Statistical boosting algorithms have their roots in machine learning \cite{Freund90} but were later adapted in order to estimate classical statistical models \cite{friedman_2001, friedmanetal2000}. A pivotal aspect of these algorithms is that they incorporate data-driven variable selection and shrinkage of effect estimates similar to the one of classical penalized regression \cite{hepp2016}. 

In a review some years ago~\cite{mayr2014evolution}, we highlighted this evolution of boosting from machine-learning to statistical modelling. Furthermore, we emphasized the similarity of two boosting approaches -- gradient boosting \cite{BuhlmannHothorn06} and likelihood-based boosting \cite{TutzBinder} --  introducing \emph{statistical boosting} as a generic term for these kind of algorithms. Throughout this article, we will use this term to reflect both approaches. 

The earlier review~\cite{mayr2014evolution} was accompanied by a second article~\cite{mayr2014extending}  highlighting the multiple variants the basic algorithms have been extended towards (i) enhanced variable selection properties, (ii) new types of predictor effects, and (iii) new regression settings. The substantial new methodological developments on statistical boosting algorithms throughout the last years (e.g., stability selection \cite{Hofner:StabSel:2015}), opening the door for the growing community to new model classes and frameworks (e.g., joint models \cite{waldmann2016boosting} and functional data \cite{brockhaus2016boosting}), make it necessary to provide an update on the available extensions. 

This article is structured as follows: In Section 2 we shortly highlight both basic structure and properties of statistical boosting algorithms and point to their connections to classical penalization approaches like the lasso. In Section 3 we focus on new developments regarding variable selection, which can be also combined with boosted functional regression models presented in Section 4. Section 5 focuses on advanced survival models before we briefly summarize in Section 6 what other relevant developments and applications have been proposed for the framework of statistical boosting.

\section{Statistical boosting}
\label{section2}

\subsection{From machine learning to statistical models}

The original boosting concept by Schapire \cite{Schapire1989} and Freund \cite{Freund90} emerged from the field of supervised learning, focusing on \textit{boosting} the accuracy of weak classifiers (\textit{base-learners}) by iteratively applying them to re-weighted data to get stronger results.  

Even if the base-learners individually only slightly outperform random guessing the combined ensemble solution can often be boosted to a perfect classification~\cite{boostingbook}. The introduction of AdaBoost \cite{nr:freund.schapire:1996} was the breakthrough for boosting in the field of supervised machine learning, allegedly leading Leo Breiman to praise its performance: \textit{Boosting is the best off-the-shelf classifier in the world} \cite{elements2}.

The main target of classical machine-learning approaches is predicting observations $y_{\text{new}}$ of the outcome $Y$ given one or more input variables $\bm{X} = \{X_1, \dots,X_p\}$. The estimation of the prediction or generalization function is based on an observed sample $(y_1, \bm{x}_1),$ $\dots,$ $(y_n, \bm{x}_n)$. However, since the underlying nature of the data generating process is treated as unknown, the focus is not on quantifying or describing this process, but solely on predicting $\hat{y}_{\text{new}}$ for new observations $x_{\text{new}}$ as accurately as possible.

As a consequence, many machine-learning approaches (also including the original AdaBoost with trees or stumps) should mainly be seen as black box prediction schemes. Although typically yielding accurate predictions \cite{wyner2015explaining}, they do not offer much insight into the structure of the relationship between explanatory variables $\bm{X}$ and the outcome $Y$.

Statistical regression models on the other hand, particularly aim at describing and explaining the underlying relationship in a structured way. The impact of single explanatory variables can not only be quantified in terms of variable importance measures \cite{strobl2007bias,hapfelmeier2014new}, but the actual effect of these variables is interpretable. The work of Friedman {\it et al.}~\cite{friedman_2001,friedmanetal2000} laid the groundwork to understand the concept of boosting from a statistical perspective and to adapt the general idea in order to estimate statistical models. 

\subsection{General model structure}

The aim of \textit{statistical boosting} algorithms is to estimate and select the effects in structured additive regression models. The most important model class are generalized additive models ('GAM', \cite{hastietib}), where the conditional distribution of the response variable is assumed to follow an exponential family distribution. Then, the expected response is modeled given the observed value $\bm{x}$ of one or more  explanatory variables using a link-function $g$ as 
\begin{equation*}
    g(\mathds{E}(Y|X=\bm{x})) =  f(\bm{x}). 
\end{equation*}

In the typical case of multiple explanatory variables,  the function $f(\cdot)$ is called additive predictor and consists of the additive effects of the single predictors,
\begin{equation}\label{eq:star}
	f(\bm{x}) = \beta_0 +  f_1(x_1) + \cdots + f_p(x_p)
\end{equation}
where $\beta_0$ represents a common intercept and the functions $f_j(x_j)$, $j = 1, \dots p$ are the partial effects of the variables $x_j$. The generic notation $f_j(x_j)$ may comprise different types of predictor effects such as classical linear effects $x_j\beta_j$, smooth non-linear effects constructed via regression splines, spatial effects or random effects of the explanatory variable $x_j$, to name but a few. 
In statistical boosting algorithms, the different partial effects are estimated by separate base-learners $h_1(\cdot), ..., h_p(\cdot)$ (\textit{component-wise boosting}, \cite{BuhlmannHothorn06}) which are typically simple regression-type prediction functions.

\begin{algorithm}[]
	\begin{enumerate}
		
		\item[] {\bf Initialization}
		\begin{enumerate}
			
			\item[(1)]  Start with iteration counter $m=0$. Initialize the additive predictor $\hat{f}^{[0]}$ with an offset value.  Specify a set of prediction functions as base-learners $h_1(x_1),..., h_p(x_p)$; typically each base-learner is a regression function incorporating one possible candidate variable.
			
		\end{enumerate}
		
		\item[]{\textbf{Component-wise fitting of base-learners}}
		
		\begin{enumerate}
			\item[(2)] Set iteration counter $m := m + 1$.
			
			\item[(3)] Fit the base-learners $\hat{h}_j(\cdot)$, $j = 1,\dots, p$, one-by-one:
			
			\paragraph{Gradient boosting}
			Base-learners are fitted to the negative gradient vector of the loss function (e.g. the log-likelihood), evaluated at the current additive predictor $\hat{f}^{[m-1]}$. 
			To ensure small steps, the base-learner fits are multiplied by a small step-length factor $\nu$, $0 \leq \nu \leq 1$:\\ $\hat{h}_j(\cdot) \ := \ \nu \cdot \hat{h}_j(\cdot) \ .$

			\paragraph{Likelihood-based boosting}
			Base-learners are estimated via maximizing the overall likelihood, using one step of Fisher scoring with the current additive predictor $\hat{f}^{[m-1]}$ as offset. To ensure small steps, a penalty term is attached to the likelihood.

		\end{enumerate}
		
		\item[]{\textbf{Update best performing component}}
		
		\begin{enumerate}
			\item[(4)] Select the best performing base-learner $j^*$:
			
			\paragraph{Gradient boosting} Based on the smallest residual sum of squares with respect to the negative gradient vector. 
			\paragraph{Likelihood-based boosting} Based on the largest overall likelihood after the update. \\

			\item[(5)] Update the additive predictor via the corresponding base-learner:
			\begin{equation*}
				\hat{f}^{[m]} = \hat{f}^{[m-1]} + \hat{h}_{j^*}(\cdot)
			\end{equation*}
			
		\end{enumerate}
		
		\item[]{\textbf{Iteration}}
		
		\begin{enumerate}
			\item[] Iterate steps (2) to (5) until $m=m_{\text{stop}}$. \\
		\end{enumerate}

	\end{enumerate}
	\caption{The structure of statistical boosting algorithms.}\label{alg:statboost}
\end{algorithm}

\subsection{Gradient boosting}\label{sec:gradient_boosting}

Gradient boosting \cite{BuhlmannHothorn06, friedman_2001} is one of the two important approaches in the context of statistical boosting. For a generic overview on the structure of statistical boosting algorithms see Box~\ref{alg:statboost}. 

In gradient boosting, the iterative procedure fits the base-learners $h_1(x_1),..., h_p(x_p)$ one-by-one to the negative gradient of the loss function $\rho(y,f(\cdot))$, evaluated at the previous iteration:
\begin{equation} 
\bm{u} =  \left(-  \left. \frac{\partial }{\partial f} \rho(y_i,f(\cdot) )\right|_{ f = \hat{f}^{[m-1]}(\cdot)} \right)_{i = 1,...,n}.
\end{equation}

The loss function describes the discrepancy between the observed outcome $y$ and the additive predictor $f(\cdot)$ and is the target function that should be minimized to get an optimal fit for $f(\cdot)$. In case of GAMs, the loss function is typically the negative log-likelihood of the corresponding exponential family. For Gaussian distributed outcomes, this reduces to the $L_2$ loss $\rho(y, f(\cdot)) = (y - f(\cdot))^2$, where the gradient vector $\bm{u}$ is simply the vector of residuals $y - f(\cdot)$ from the previous iteration and boosting hence corresponds to refitting of residuals. 

In each boosting iteration, only the best-fitting base-learner $h_{j^*}$ is selected based on the residual sum of squares of the base-learner fit
\begin{equation} 
j^* = \underset{1 \leq j \leq p}{\operatorname{argmin}}\sum_{i=1}^n (u_{i} - \hat{h}_{j}(x_j))^2 \ . 
\end{equation}
Only this base-learner $h_{j^*}$ is added to the current additive predictor $f(\cdot)$. In order to ensure small updates, only a small proportion of the base-learner fit (typically the step length is $\nu = 0.1$ \cite{BuhlmannHothorn06}) is actually added. Note that the base-learner $h_j(\cdot)$ can be selected and updated various times; the partial effect of variable $x_j$ is the sum of all corresponding base-learner that had been selected: 
\begin{equation*}
\hat{f}_j(x_j) = \sum_m  \nu \cdot  \hat{h}_j(x_j) \mathds{I}_{j = j*} \ . 
\end{equation*}

This component-wise procedure of fitting the base-learners one by one to the current gradient of the loss function can be described as \textit{gradient descent in function space} \cite{mason1999boosting}, where the function space is spanned by the base-learners. The algorithm effectively optimizes the loss function step-by-step, eventually converging to the minimum. In order to avoid overfitting and to ensure variable selection, the algorithm is typically stopped before convergence (based on predictive performance evaluated via cross-validation or resampling \cite{Mayr_mstop}), which leads to an implicit penalization \cite{hothorn2014boosting}. 

Gradient boosting is implemented in the add-on package \textbf{mboost} \cite{pkg:mboost:CRAN} for the  open source programming environment \textsf{R} \cite{r-core:2016}, providing a large number of pre-implemented loss functions for various regression settings, as well as different base-learners to represent various types of effects (see \cite{mboostTut} for an overview). Recent changes in the software, which were introduced after the comprehensive mboost tutorial \cite{mboostTut} are provided as Appendix~\ref{seq:mboost_changes}.

\subsection{Likelihood-based boosting}

Likelihood-based boosting \cite{TutzBinder, TutBin2007} is the other general approach in the framework of statistical boosting algorithms. It follows a very similar structure as gradient boosting (see Box~\ref{alg:statboost}), although both approaches only coincide in special cases such as classical Gaussian regression via the $L_2$ loss \cite{mayr2014evolution,BuehlmannYu2003}. In contrast to gradient boosting, the base-learners are directly estimated via optimizing the overall likelihood, using the additive predictor from the previous iteration as offset. In case of the $L_2$ loss, this has a similar effect than refitting the residuals.

In every step, the algorithm hence optimizes regression models as base-learners one-by-one by maximizing the likelihood (using one step Fisher scoring), selecting only the base-learner $j^*$ which leads to the largest increase in the likelihood. In order to obtain small boosting steps, a quadratic penalty term is attached to this likelihood. This has a similar effect as multiplying the fitted base-learner by a small step-length factor as in gradient boosting.

Likelihood-based boosting for generalized linear and additive regression models is provided by the \textsf{R} add-on package \textbf{GAMBoost} \cite{GAMBoost}, and an adapted version for boosting Cox regression is provided with \textbf{CoxBoost} \cite{CoxBoost}. For a comparison of both statistical boosting approaches, i.e., likelihood-based and gradient boosting in case of Cox proportional hazard models, we refer to \cite{deBin2016boosting}. 

\subsection{Connections to $L_1$-regularization}

Statistical boosting algorithms result in regularized models with shrinked effect estimates although they only apply implicit penalization \cite{hothorn2014boosting} by stopping the algorithm before convergence. By performing regularization without the use of an explicit penalty term, boosting algorithms clearly differ from other direct regularization techniques such as the \emph{lasso} \cite{Tibsh96}. However, both approaches sometimes result in very similar models after being tuned to a comparable degree of regularization~\cite{hepp2016}. 

This close connection has been first noted between the lasso and \emph{forward stagewise regression}, which can be viewed as special case of the gradient boosting algorithm (Box \ref{alg:statboost}), and led, along with the development of \emph{least angle regression} (LARS), to the formulation of the \emph{positive cone condition} (PCC) \cite{efronetal}.

If this condition holds, LARS, lasso and forward stagewise regression coincide.
Figuratively speaking, the PCC requires that all coefficient estimates monotonically increase or decrease with relaxing degree of regularization and applies, for example, to the case of low-dimensional settings with orthogonal $X$. 
It should be noted that the PCC is connected to the \emph{diagonal dominance condition} for the inverse covariance matrix of $X$, which allows for a more convenient way to investigate the equivalence of these approaches in practice \cite{meinshausen2007threecousins,duan2012conditions}.

Given that the solution of the lasso is optimal with respect to the $L_1$-norm of the coefficient vector, these findings led to the notion of boosting as some ``sort of $L_1$-sparse'' regularization technique \cite{buhlmann2014discussion}, but it remained unclear which optimality constraints possibly apply to forward stagewise regression if the PCC is violated.

By further extending $X$ with a negative version of each variable and enforcing only positive updates in each iteration, Hastie {\it et al.}~\cite{hastie2007monotone} demonstrated that forward stagewise regression always approximates the solution path of a similarly modified version of the lasso.
From this perspective, they showed that forward stagewise regression minimizes the loss function subject to the $L_1$\emph{-arc-length}
\begin{equation*}
 \sum_{j=1}^p \int^t_0 \left| \frac{\partial\beta_j(s)}{\partial s}\right| ds \le t. 
\end{equation*}

This means that the travelled path of the coefficients is penalized (allowing as little overall changes in the coefficients as possible, regardless of their direction), whereas the $L_1$-norm considers only the absolute sum of the current set of estimates. In the same article, Hastie {\it et al.}~\cite{hastie2007monotone} further showed that these properties hold for general convex loss functions and therefore apply not only to forward stagewise regression but for the more general gradient boosting method (in case of logistic regression models as well as for many other generalized linear regression settings). 

The consequence of these differing optimization constraints can be observed in the presence of strong collinearity, where the lasso estimates tend to be very unstable regarding different degrees of regularization while boosting approaches avoid too many changes in the coefficients as they consider the overall travelled path \cite{hepp2016}.

It has to be acknowledged, however, that direct regularization approaches as the lasso are applied more often in practice \cite{buhlmann2014discussion}. Statistical boosting, on the other hand, is far more flexible due to its modular nature allowing to combine any base-learner with any type of loss-function \cite{hepp2016,buhlmann2014discussion}. 

\section{Enhanced variable selection}
\label{section3}

Early stopping of statistical boosting algorithms via  cross-validation approaches plays a vital role to ensure a sparse model with optimal prediction performance on new data. Resampling, i.e., random sampling of the data drawn without replacement, tends to result in sparser models compared to other sampling schemes \cite{Mayr_mstop}, including the popular bootstrap \cite{janitza2015pitfalls}. By using base-learners of comparable complexity (in terms of degrees of freedom) selection bias can be strongly reduced \cite{Hofner:unbiased:2011}. The resulting models have optimal prediction accuracy on the test data. Yet, despite regularization the final models are often relatively rich \cite{Mayr_mstop}. 

\subsection{Stability selection}\label{sec:stabsel}
Meinshausen and Bühlmann \cite{Meinshausen:2008} proposed a generic 
approach called stability selection to further refine the models and enhance sparsity. 
This approach was then transferred to boosting \cite{Hofner:StabSel:2015}. 
In general, stability selection can be combined with any variable selection approach 
and is especially useful for high-dimensional data with many potential predictors. To
assess how stable the selection of a variable is, $B$ random subsets that comprise half of the data are drawn. On each of these subsets, the model is fitted until $q$ base-learners are  selected. Usually, $B = 100$ subsets are sufficient. Computing the relative frequencies of random subsamples in which specific base-learners were  selected give a notion of how stable the selection is with respect to perturbations of the data. Base-learners are considered to be of importance if the selection frequency exceeds a pre-specified threshold level $\pi_{\text{thr}} \in [0.5, 1]$. 

Meinshausen and Bühlmann \cite{Meinshausen:2008} showed that this approach controls 
the per-family error rate (PFER), i.e., it provides an upper bound for the expected number of false positive selections ($V$):
\begin{equation}\label{eq:pfer}
  \mathds{E}(V) \leq \frac{q^2}{(2\pi_{\text{thr}} - 1) p},
\end{equation}
where $p$ is the number of base-learners. This upper bound is rather conservative
and hence was further refined by Shah and Samworth \cite{shah2013variable} for
specific assumptions on the distribution of the selection frequencies. Stability
selection with all available error bounds is implemented for a variety of modelling
techniques in the \textsf{R} package \textbf{stabs} \cite{Hofner:pkg_stabs:2017}.

An important issue is the choice of the hyper-parameters of stability selection. 
The choice of a fix value of $q$ should be made such that it is large enough 
to select all hypothetically influential variables \cite{Hofner:StabSel:2015,mayr2016boosting}. A sensible value for $q$ should
usually be smaller or equal to the number of base-learners selected via early stopping with cross-validation. 

In general, the size of $q$ is of minor importance if it is in a sensible range. 
With a fixed $q$ the threshold for 
stable effects either the threshold for $\pi_{\text{thr}}$ can be chosen additionally or, as can be seen from Equation~\eqref{eq:pfer}
using equality, the upper bound for the PFER can be pre-specified and 
the threshold can be derived accordingly. The latter would be the preferred choice if error 
control is of major importance, the former if error control is just considered
a by-product (see e.g., \cite{mayr2016boosting}). For an interpretation of the 
PFER, particularly with regard to standard error rates such as the per-comparison
error rate or the family-wise error rate, we refer to Hofner {\it et al.}~\cite{Hofner:StabSel:2015}. 
Note that for a fixed $q$, it is computationally very easy to change any of the 
other two parameters ($\pi_{\text{thr}}$ or the upper bound for the PFER) as the 
resampling results can be reused \cite{Hofner:StabSel:2015}.

Please note that base-learners selected via stability selection might not reflect any model which can be derived with a specific penalty parameter using the original modelling approach. This means that for boosting, no $m_{\text{stop}}$ value might exist that results in a model with the stably selected base-learners; the provided set of stable base-learners is a fundamentally new solution. 

\subsection{Extension and application of boosting with stability selection}\label{sec:selection_extension}

Variable selection is especially important in high-dimensional gene expression data and other large scale biomedical data sources. 
Recently, stability selection with boosting was successfully applied to select a small number of informative biomarkers for survival of breast cancer patients \cite{mayr2016boosting}.
The model was derived based on a novel boosting approach that optimizes the concordance index \cite{mayr2014CI,chen2013gradient}.
Hence, the resulting prediction rule was optimal with respect to its ability to discriminate between patients with longer and shorter survival, i.e., its discriminatory power. 

Thomas {\it et al.}~\cite{thomas2016stability} derived a modified algorithm for boosted
generalized additive models for location, scale and shape (GAMLSS, \cite{rs}) to allow a combination of this very flexible model class with stability selection. 
The basic idea of GAMLSS is to model all parameters of the conditional distribution by their own additive predictor and associated link function.
Extensive simulation studies showed that the new fitting algorithm leads to comparable models as the previous algorithm \cite{gamboostlss:2012,Hofner:gamboostLSS_Tutorial:2016} but is superior regarding the computational speed, especially in combination with cross-validation approaches. Furthermore, simulations showed that this algorithm can be successfully combined with stability selection to select sparser models identifying a smaller subset of truly informative variables from high-dimensional data. The current algorithm is implemented in the \textsf{R} add-on package \textbf{gamboostLSS}~\cite{hofner:pkg:gamboostlss}, the modified version is currently available on GitHub~\cite{hofner:gamboostlss:github}.

\subsection{Further approaches for sparse models}

In order to construct risk prediction signatures on molecular data, such as DNA methylation, Sariyar {\it et al.}~\cite{sariyar2014boosting} proposed an adaptive likelihood-based boosting algorithm. The authors included a step size modification factor $c_f$ which represents an additional tuning parameter, adaptively controlling the size of the updates. In case of sparse settings, the approach decreases shrinkage of effect estimates (by using a larger step-length) leading to a smaller bias. In settings with larger numbers of informative variables, the approach allows to fit models with lower degree of sparsity when necessary by smaller updates. The modification factor $c_f$ has to be selected together with $m_{\text{stop}}$ via cross-validation or resampling on a two-dimensional grid. 

Zhang {\it et al.}~\cite{zhang2016pboostga} argue that variable ranking in practice is more favourable than variable selection, as ranking allows to easily apply a thresholding rule in order to identify a subset of informative variables. The authors implemented a pseudo-boosting approach, which is technically not based on statistical boosting but is adapted to rank and select variables for statistical models.
Note that also stability selection can be seen as a variable ranking scheme based on their selection frequency, as its selection feature is only triggered by implementing the threshold $\pi_{\text{thr}}$.

Following a gradient based approach, Huang {\it et al.}~\cite{huang2017promoting} adapted the sparse boosting approach by Bühlmann and Yu~\cite{buhlmann2006sparse} in order to promote similarity of model sparsity structures in the integrative analysis of multiple data sets, which surely is an important topic regarding the trend toward big data.

\section{Functional regression}

Due to technological developments, more and more data is measured continuously over time. Over the last years, a lot of methodological research focused on regression methods for this type of functional data. A groundbreaking work in this new and evolving field of statistics is provided by Ramsay and Silverman \cite{james2005functional}. 

Functional regression models can either contain functional responses (defined on a continuous domain), functional covariates or both. This leads basically to three different classes of functional regression models, i.e., function-on-scalar (response is functional), scalar-on-function (functional explanatory variable) and function-on-function regression. For a recent review on general methodological developments on functional regression, see Morris \cite{morris2015functional}.

\subsection{Boosting functional data}

The first statistical boosting algorithm for functional regression, allowing for data-driven variable selection, was proposed by Brockhaus {\it et al.}~\cite{brockhaus2015functional}. The authors' approach focused on linear array models \cite{currie2006generalized} providing a unified framework for all three settings outlined above. Since the general structure of their gradient boosting algorithm is similar to the one in Box~\ref{alg:statboost}, the resulting models still have the same form as in (\ref{eq:star}), only that the response $Y$ and the covariates may be functions. The underlying functional partial effects $h_{j}(x_j, t)$ can be represented using tensor product basis 
\begin{equation*}
	h_j(x_j)(t) = \left(b_j(x_j)^{\top} \otimes b_Y(t)^{\top} \right)\theta_j  \ ,
\end{equation*}
where $\theta_j$ is the vector of coefficients, $b_j$ and $b_Y$ are basis functions, and $\otimes$ denotes the Kronecker product. 

This functional array model is limited in two ways: (i) the functional responses need to be measured on a common grid and (ii) covariates need to be constant over the domain of the response.  As particularly the second assumption might often not be fulfilled in practice, Brockhaus {\it et al.}~\cite{brockhaus2016boosting} soon after proposed a general framework for boosting functional regression models avoiding this assumption and dropping the linear array structure.  

This newer framework \cite{brockhaus2016boosting} comprises also all three model classes outlined above and particularly focuses on historical effects, where functional response and functional covariates are observed over the same time interval. The underlying assumption is that observations of the covariate affect the response only up to the corresponding time point $t$
\begin{equation}\label{eq:histfun}
\mathds{E}(Y(t) | X = x) = \sum_{j = 1}^J \int_{t_1}^{t} x_j(s) \beta_j(s,t) ds \ ,
\end{equation}
where $s$ represents the time points the covariate was observed for.
In other words, only the part of the covariate function lying in the past (not the future) can affect the present response.
However, this is a sensible restriction in most practical applications and thus not a strong restriction.

Both approaches for boosting functional regression are implemented in the \textsf{R} add-on package \textbf{FDboost} \cite{FDboost:CRAN}, which relies on the fitting methods and infrastructure of \textbf{mboost}.

\subsection{Extensions of boosting functional regression}

Boosting functional data can be combined with stability selection (see Section~\ref{sec:stabsel}) in order to enhance the variable selection properties of the algorithm \cite{brockhaus2015functional,brockhaus2016boosting}.

The boosting approach for functional data was already extended towards the model class of generalized additive models for location, scale and shape (GAMLSS) for a scalar-on-function setting by Brockhaus {\it et al.}~\cite{brockhaus2016signal}.  The functional approach was named signal regression models for location, scale and shape \cite{brockhaus2016signal}. The estimation via gradient boosting is based on the corresponding gamboostLSS algorithm for boosting GAMLSS \cite{gamboostlss:2012,Hofner:gamboostLSS_Tutorial:2016}. 

In an approach to analyse the functional relationship between bioelectrical signals like electroencephalography (EEG) and facial electromyography (EMG), Rügamer {\it et al.}~\cite{rugamer2016boosting} focused on extending the framework of boosting functional regression by incorporating factor specific historical effects, similar to (\ref{eq:histfun}).

Although functional data analysis triggered a lot of methodological research, a recent systematic review by Ullah and Finch \cite{ullah2013applications} revealed that the number of actual biomedical applications of functional data analysis in general and functional regression in particular is rather small. The authors argued that the potential benefits of these flexible models (like richer interpretation and more flexible structures) are not yet well understood by practitioners and that further efforts are necessary to promote the actual usage of these novel techniques. 

\section{Boosting advanced survival models}

While Cox regression is still the dominant model class for boosting time-to-event data (see \cite{deBin2016boosting} for a comparison of two different boosting algorithms, and \cite{zemmour2015prediction} for different general approaches to estimate Cox models in the presence of high-dimensional data), over the last years several alternatives emerged \cite{mayr2014CI,chen2013gradient,Schmid:Hothorn:AFT-boost}.  

In this section we will particularly focus on boosting joint models of time-to-event outcomes and longitudinal markers  but will also briefly refer to other recent extensions.

\subsection{Boosting joint models}\label{sec:JMboost}

The concept of joint modelling of longitudinal and time-to-event data has found its way into the statistical literature in the last few years as it gives a very complete answer to questions on continous data recorded over time and event times related to this continous data. Modelling those to processes independently as done up to the suggestion of the joint modelling idea~\cite{wulfsohn1997} leads to misspecified models prone to bias. There are various joint modelling approaches and thus also various different model equations. The type we are going to refer to in this review are of the following type:
\begin{eqnarray}
 y_{ij} &= &\eta_{\text{l}}(x_{ij}) + \eta_{\text{ls}}(x_i,t_{ij}) + \varepsilon_{ij}\nonumber\\
 \lambda(t|\alpha,\eta_{\text{s}}(x_i, t),\eta_{\text{ls}}(x_i, t))& =& \lambda_0(t)\exp(\eta_{\text{s}}(x_i, t) + \alpha\eta_{\text{ls}}(x_i, t) ),
\label{jm_like}
\end{eqnarray}
where $y_{ij}$ is the $j$-th observation of the $i$-th individual with $i = 1,\ldots,n$ and $j=1,\ldots,n_i$ and $\lambda(t|\alpha,\eta_{\text{s}}(x_i, t),\eta_{\text{ls}}(x_i, t))$ is the hazard function for individual $i$ at time point $t$. Both outcomes, the longitudinal measurement as well as the event time are modeled based on two sub-predictors each: one that is supposed to have an impact on only one of them (the longitudinal sub-predictor $\eta_{\text{l}}(x_{ij})$ and the survival sub-predictor $\eta_{\text{s}}(x_{ij},t)$) and one of them being shared by both parts of the model (the shared sub-predictor $\eta_{\text{ls}}(x_{ij},t)$). All those sub-predictors are functions of different, possibly time dependent variables $x_i$. In many cases the shared sub-predictor consists of or at least includes some type of random effects. The function $\lambda_0(t)$ is the baseline hazard. Most approaches for joint models are based on likelihood or Bayesian inference using the joint likelihood resulting as a product from the above likelihoods \cite{faucett1996,JM}. Those approaches are, however, unable to conduct variable selection and cannot deal with high-dimensional data. 

Waldmann {\it et al.}~\cite{waldmann2016boosting} suggested a boosting algorithm tackling these challenges. The model used in this paper was a reduced version of~\eqref{jm_like} in which no survival sub-predictor was considered and a fixed baseline hazard $\lambda_0$ was used. The algorithm is a version of the classical boosting algorithm as represented in Box~\ref{alg:statboost}, which is adapted to the special case of having to estimate a set of different sub-predictors (similar to \cite{gamboostlss:2012}). The algorithm is therefore composed of three steps which are performed circularly.
In the first step a regular boosting step to update the longitudinal sub-predictor $\eta_{\text{l}}(x_{ij})$ is performed and the parameters of the shared sub-predictor are treated as fixed.
In the second step, the parameters of the longitudinal sub-predictor are fixed and a boosting step for the shared sub-predictor $\eta_{\text{ls}}(x_{ij})$ is conducted.
The third step is a simple optimization step: based on the current values of the parameters in both sub-predictors the likelihoods are optimized with respect to $\lambda_0$, $\sigma^2$ and $\alpha$ (cf., \cite{Schmid:Multidim:2010}). The number of iterations now depends on two stopping iterations $m_{\text{stop, l}}$ and $m_{\text{stop, ls}}$ which have to be optimized on a two dimensional grid via cross validation. Waldmann {\it et al.}~\cite{waldmann2016boosting} showed that the benefits of boosting algorithm (automated variable selection and handling of $p>n$ situations) can be transfered to joint modelling and hence lay the groundwork to further extended joint modelling approaches.

The code for the approach presented here is available in the \textsf{R} add-on package \textbf{JMboost}~\cite{jmboost}, currently on GitHub.

\subsection{Other new approaches on boosting survival data}

Reulen and Kneib \cite{reulen2016boosting} extended the framework of statistical boosting towards multi-state models for patients exposed to competing risks (e.g., adverse events, recovery, death or relapse). The approach is implemented in the \textbf{gamboostMSM} package \cite{gamboostMSM},  relying on the infrastructure of \textbf{mboost}.
M\"ost and Hothorn~\cite{most2015conditional} focused on boosting patient-specific survivor function based on conditional transformation models \cite{ctm} incorporating inverse probability of censoring weights \cite{van2003unified}.

When statistical boosting algorithms are used to estimate survival models, the motivation is most often the presence of high-dimensional data. De Bin {\it et al.}~\cite{de2014investigating} investigated several approaches (including gradient boosting and likelihood-based boosting) to incorporate both clinical and high-dimensional omics data to build prediction models. 

Guo {\it et al.}~\cite{guo2015forward} proposed a new adaptive likelihood-based boosting algorithm to fit Cox models, incorporating a direct lasso-type $L_1$ penalization in the fitting process in order to avoid the inclusion of variables with small effect. This general motivation is similar to the one of the boosting algorithm with step-length modification factor proposed by Sariyar {\it et al.}~\cite{sariyar2014boosting}. In another approach, Sariyar {\it et al.}~\cite{sariyar2014combining} combined a likelihood-based boosting approach for the Cox model with random forest in order to screen for interaction effects in high-dimensional data. Hieke {\it et al.}~\cite{hieke2016identifying} combined likelihood-based boosting with resampling in an approach to identify prognostic SNPs in potentially small clinical cohorts.

\section{New frontiers and applications}

There were even more new topics that have been incorporated into the framework of statistical boosting, but not all of them can be presented in detail here. However, we want to give a short overview of the most relevant developments, notably many of those were actually motivated by biomedical applications. 

Weinhold {\it et al.}~\cite{weinhold2016statistical} proposed to analyse DNA methylation data (signal intensities $M$ and $U$), via a ``ratio of correlated gammas'' model. Based on a bivariate gamma distribution for $M$ and $U$ values, the authors derived the density for the ratio $\frac{M}{M+U}$ and optimized it via gradient boosting.

A boosting algortihm for differential item functioning in Rasch models was developed by  Schauberger and Tutz~\cite{schauberger2016detection} for the broader area of psychometrics, while Casalicchio {\it et al.} focused on a boosting subject-specific Bradley-Terry-Luce models \cite{casalicchio2015subject}. 

Napolitano {\it et al.}~\cite{napolitano2017predicting} developed a sampled boosting algorithm for the analysis of brain perfusion images: Gradient boosting is carried out multiple times on different training sets. Each base-learner refers to a voxel and after every sampling iteration a fixed fraction of selected voxels is randomly left out from the following boosting fit, in order to force the algorithm to select new voxels. The final model is then computed as the global sum of all solutions. Feilke {\it et al.}~\cite{feilke2016boosting} proposed a voxelwise boosting approach for the analysis of contrast-enhanced magnetic resonance imaging data (DCE-MRI), which was additionally enhanced to account for the regional structure of the voxels via a spatial penalty. 

Pybus {\it et al.}~\cite{pybus2015hierarchical} proposed a hierarchical boosting algorithm for classification in an approach to detect positive selection in genomic regions (cf., \cite{lin2011distinguishing}). Truntzer {\it et al.}~\cite{truntzer2014comparison} compared the classification performance of gradient boosting with other methods combining clinical variables and high-dimensional mass spectrometry data and concluded that the variable selection properties of boosting led also to a very good performance regarding prediction accuracy. 

Regarding boosting location and scale models (modelling both expected value and variance in the spirit of GAMLSS \cite{rs}), Messner {\it et al.}~\cite{messner2017nonhomogeneous} proposed a boosting algorithm for predictor selection in ensemble postprocessing to better calibrate ensemble weather forecasts. The idea of ensemble forecasting is to account for model errors and to quantify forecast uncertainty. Mayr {\it et al.}~\cite{mayr2015permutation} used boosted location and scale models in combination with permutation tests to assess simultaneously systematic bias and random measurement errors of medical devices. The use of a permutation test tackles one of the remaining problems of statistical boosting approaches in practical biomedical research: The lack of standard errors for effect estimates makes it necessary to incorporate resampling procedures to construct confidence intervals or to assess significance of effects. 

The methodological development in \cite{mayr2015permutation}, analogously to many of the extensions presented in this article, was motivated by the applied analysis of biomedical data. Statistical boosting algorithms, however, have been applied over the last few years in various biomedical applications without the need for methodological extensions that could be described here. Most application focus on prediction modelling or variable selection. We want to briefly mention a selection of the most recent ones from the last two years: The different research questions comprise the development of birth weight prediction formulas \cite{faschingbauer2016new} for particularly small babies, the prediction of smoking cessation and its relapse in HIV-infected patients \cite{schafer2015predicting}, Escherichia coli Fed-Batch Fermentation Modeling \cite{melcher2017boosted}, the prediction of cardiovascular death on older patients in the emergency department \cite{bahrmann2015prognostic} and the identification of factors influencing therapeutic decisions regarding rheumatoid arthritis \cite{Pattloch2016}.

\section{Discussion}

After Friedman {\it et al.}~\cite{friedmanetal2000} discovered the link between boosting and additive modelling in their seminal paper, most research on boosting methods has been focused on the development of methodology within the univariate GAM framework. This line of research included, among many other achievements, the estimation of smooth  predictor effects via spline base-learners \cite{BuehlmannYu2003} and the extension of boosting  to other GAM families than binary classification and Gaussian regression \cite{BuhlmannHothorn06}. We have summarized these methods and described their relationships in an earlier review~\cite{mayr2014evolution}.

In this article, we have highlighted several new research areas in the field of statistical boosting leaving the traditional GAM modeling approach. A particularly active research area during the last few years addresses the development of boosting algorithms for new model classes extending the GAM framework. These include, among others, the simultaneous modelling of location, scale and shape parameters within the GAMLSS framework \cite{gamboostlss:2012}, the modelling of functional data \cite{brockhaus2015functional}, and, recently, the class of joint models for longitudinal  and survival data \cite{waldmann2016boosting}. It goes without saying that these developments will make boosting algorithms available for practical use in much more sophisticated clinical and epidemiological applications than before.

Another line of research, which we described in detail in Sections \ref{section2} and \ref{section3}, aims at exploring the connections between statistical boosting methods and machine learning techniques that were originally developed independently of boosting. An important example is stability selection, a generic methodology that, at the time of its development, mainly focussed on penalized regression models such as the lasso. Only in recent years, stability selection has been adapted to become a tool for variable selection within the boosting framework (e.g.\@ \cite{thomas2016stability}). Other work in this context is the analysis of the connections between boosting and penalized regression \cite{hepp2016} and the work by Sariyar {\it et al.}~\cite{sariyar2014combining} exploring a combination of boosting and random forest methods.

Finally, as already noted by Hothorn~\cite{hothorn2014boosting}, boosting may not solely be regarded as a framework for regularized model fitting but also as a generic optimization tool on its own right. In particular, boosting constitutes a robust algorithm for the optimization of objective functions that, due to their structure or complexity, may pose problems for Newson-Raphson-type and related methods. This was, for example, the motivation for the use of boosting in the articles by Hothorn {\it et al.}~\cite{ctm} and Weinhold~{\it et al.}~\cite{weinhold2016statistical}.

Regarding future research, a huge challenge for the use of boosting algorithms in biomedical applications arises from the \textit{era of big data}. Unlike other machine learning methods like random forests, the sequential nature of boosting methods hampers the use of parallelization techniques within the algorithm, which may result in issues with the fitting and tuning of complex models with multidimensional predictors and/or sophisticated base-learners like splines or higher-sized trees. To overcome these problems in classification and univariate regression, Chen and Guestrin~\cite{chen2016xgboost} developed the extremely fast and sophisticated \textbf{xgboost} environment. However, for the more recent extensions discussed in this paper, \textit{big data} solutions for statistical boosting still have to be developed.

\subsection*{Conflict of interests}

The authors declare that there is no conflict of interest regarding the publication of this
paper.

\subsection*{Acknowledgements}

The authors thank Corinna Buchstaller for her help with the literature search. The work on this article was supported by the Deutsche Forschungsgemeinschaft (DFG) (\texttt{www.dfg.de}), grant no.~SCHM 2966/1-2 (grant to MS and OG) and the Interdisciplinary Center for Clinical Research (IZKF) of the Friedrich-Alexander-University Erlangen-N\"urnberg via the Projects J49 (grant to AM) and J61 (grant to EW).

\small
\bibliography{bibliography_boosting_AM}

\newpage
\begin{appendix}
\section{Developments regarding the mboost package} \label{seq:mboost_changes}

This appendix describes important changes during the last years that were implemented in the \textsf{R} package \textbf{mboost} after the tutorial paper \cite{mboostTut} on its use was published. 

Starting from \textbf{mboost 2.2}, the default for the degrees of freedom was changed, they are now defined as 
\begin{equation*}
\mathrm{df}(\lambda) = \mathrm{trace}(2S - S^{\top}S),
\end{equation*}
with smoother matrix $S = X(X^{\top}X + \lambda K)^{-1} X$. Analyses have shown, that  this leads to a reduced selection bias, see~\cite{Hofner:unbiased:2011}. Earlier versions used the trace of the smoother matrix as degrees of freedom, i.e., 
$\mathrm{df}(\lambda) = \mathrm{trace}(S)$. One can change to the old definition 
by setting \texttt{options(mboost\_dftraceS = TRUE)}. For parallel computations of cross-validated stopping values, \textbf{mboost} now uses the package \textbf{parallel}, which is included in the standard \textsf{R} installation. The behavior of \texttt{bols(x, intercept = FALSE)} was changed when \texttt{x} is a factor: the intercept is simply dropped from the design matrix and the coding can be specified as usual for factors. Additionally, a new contrast was introduced: \texttt{"contr.dummy"} (see the manual of \texttt{bols} for details). Finally, the computation of B-spline basis at the boundaries was changed such that equidistant boundary knots are used per default.

With \textbf{mboost 2.3}, constrained effects  \cite{Hofner:monotonic:2011,Hofner:constrained:2014}
are fitted per default using quadratic programming methods (option \texttt{type = "quad.prog"}) 
improving the speed of computation drastically. Additional to monotonic, convex and concave
effects, new constraints were introduced to fit \texttt{"positive"} or \texttt{"negative"} effects or effects with boundary constraints (see \texttt{bmono} for details). Additionally, a new function to assign $m_{\text{stop}}$ values to a model 
object was added (\texttt{mstop(mod) <- i}) as well as  two new distribution families \texttt{Hurdle} \cite{hofner2015seabird} and \texttt{Multinomial} \cite{Schmid:Multidim:2010}. Finally, a new option was implemented to allow for stopping based on out-of-bag data during fitting (via 
\texttt{boost\_control(..., stopintern = TRUE)}).

With \textbf{mboost 2.4}, bootstrap confidence intervals were implemented in the 
novel \texttt{confint} function \cite{Hofner:constrained:2014}. The stability
selection procedure was moved to a dedicated package \textbf{stabs}~\cite{Hofner:pkg_stabs:2017}, 
while a specific function for gradient boosting was implemented in package 
\textbf{mboost}.

From \textbf{mboost 2.5} onward, cross-validation does not stop on errors in single 
folds anymore and was sped up by setting \texttt{mc.preschedule = FALSE} if parallel
computations via \texttt{mclapply} are used. A documentation for the function
\texttt{plot.mboost} was added, which allows to visualize model results. Values outside
the boundary knots are now forbidden during fitting, while linear extrapolation 
is used for prediction. 

With \textbf{mboost 2.6} a lot of bug fixes and small improvements were provided. 
Most notably, the development of the package is now hosted entirely on github in 
the collaborative project \href{https://github.com/boost-R/mboost/}{boost-R/mboost} 
and the package maintainer changed.

The current CRAN version \textbf{mboost 2.7} provides a new family 
\texttt{Cindex}~\cite{mayr2014CI}, variable importance measures (\texttt{varimp}) 
and improved plotting facilities.

Changes in the current development version which will be deployed to CRAN with the
next release of \textbf{mboost} include major changes to distribution families, allowing to
specify link functions. The \texttt{Binomial} family will additionally
provide an alternative implementation of Binomial regression models along the lines 
of the classic \texttt{glm} implementation, which can be used via \texttt{Binomial(type = "glm")}. 
This family also works with a two-column matrix containing the number of successes 
and number of failures. Furthermore, models with zero steps (i.e., models containing 
only the offset) will be supported as well as cross-validated models without 
base-learners.

\end{appendix}

\end{document}